\documentclass{jkas}

\def\year{2016} 
\def\month{June} 
\def\volume{49} 
\def\issue{3} 
\def\beginpage{73} 
\def\endpage{81} 
\def\received{February 27, 2016} 
\def\accepted{April 27, 2016} 

\date{Received \received ; accepted \accepted}

\usepackage{flushend} 

\title{
A Super-Jupiter Microlens Planet Characterized by High-Cadence KMTNet Microlensing Survey Observations\\ of OGLE-2015-BLG-0954
}

\author[1]{I.-G.~Shin}
\author[2]{Y.-H.~Ryu}
\author[3]{A.~Udalski}
\author[4]{M.~Albrow}
\author[2,5]{S.-M.~Cha}
\author[6]{J.-Y.~Choi}
\author[2]{S.-J.~Chung}
\author[7]{C.~Han}
\author[7]{K.-H.~Hwang}
\author[1]{Y.~K.~Jung}
\author[2]{D.-J.~Kim}
\author[2,8]{S.-L.~Kim}
\author[2,8]{C.-U.~Lee}
\author[2,5]{Y.~Lee}
\author[2,8]{B.-G.~Park}
\author[6]{H.~Park}
\author[9]{R.~W.~Pogge}
\author[1,10]{J.~C.~Yee}
\author[3]{P.~Pietrukowicz}
\author[3]{P.~Mr{\'o}z}
\author[3]{S.~Koz{\l}owski}
\author[3,9]{R.~Poleski}
\author[3]{J.~Skowron}
\author[3]{I.~Soszy{\'n}ski}
\author[3]{M.~K.~Szyma{\'n}ski}
\author[3]{K.~Ulaczyk}
\author[3]{{\L}.~Wyrzykowski}
\author[3]{M.~Pawlak}
\author[2,9,11]{A.~Gould}

\affil[1]{Harvard-Smithsonian Center for Astrophysics, 60 Garden St., Cambridge, MA 02138, USA \email{in-gu.shin,yeonkil.jung,jyee@cfa.harvard.edu}}
\affil[2]{Korea Astronomy and Space Science Institute, 776 Daedeokdae-ro, Yuseong-gu, Daejeon 34055, Korea \email{yoonhyunryu@gmail.com,chasm,sjchung,keaton03,slkim,leecu,yslee,bgpark@kasi.re.kr,}}
\affil[3]{Warsaw University Observatory, A1. Ujazdowski 4, 00-478 Warszawa, Poland; \email{udalski@astrouw.edu.pl}}
\affil[4]{Department of Physics and Astronomy, University of Canterbury, Private Bag 4800 Christchurch, New Zealand \email{michael.albrow@canterbury.ac.nz}}
\affil[5]{School of Space Research, Kyung Hee University, Giheung-gu, Yongin, Gyeonggi-do 17104, Korea}
\affil[6]{Busan National Science Museum, Busan 46081, Korea; \email{quff176@gmail.com,crusader713@hanmail.net}}
\affil[7]{Department of Physics, Chungbuk National University, Seowon-Gu, Cheongju 28644, Korea \email{cheongho@astroph.chungbuk.ac.kr,kyuha1@gmail.com}}
\affil[8]{Korea University of Science and Technology, 217 Gajeong-ro, Yuseong-gu, Daejeon 34113, Korea}
\affil[9]{Department of Astronomy Ohio State University,
140 W.\ 18th Ave., Columbus, OH 43210, USA \email{pogge@astronomy.ohio-state.edu}}
\affil[10]{Sagan Fellow}
\affil[11]{Max-Planck-Institute for Astronomy, K\"onigstuhl 17,
69117 Heidelberg, Germany; \email{gould@astronomy.ohio-state.edu}}

\def\corrauthor{
A. Gould
}

\def\runningauthor{
Shin et al.
}

\def\runningtitle{
A Super-Jupiter Microlens Planet Characterized by KMTNet Observations
}

\def\keywords{
gravitational microlensing --- planets 
}

\def\abstracttext{
We report the characterization of a massive
$(m_p=3.9\pm 1.4 M_{\rm jup})$ microlensing planet (OGLE-2015-BLG-0954Lb) orbiting an M dwarf host
($M=0.33\,\pm 0.12 M_\odot$) at a distance toward the Galactic bulge
of $0.6^{+0.4}_{-0.2}\,$kpc, which is extremely nearby by microlensing
standards.  The planet-host projected separation is
$a_\perp \sim 1.2\,\au$.  The characterization was made possible
by the wide-field ($4\,\rm deg^2$) high cadence ($\Gamma = 6\,\rm hr^{-1}$)
monitoring of the Korea Microlensing Telescope Network (KMTNet),
which had two of
its three telescopes in commissioning operations at the time of the
planetary anomaly.  The source crossing time $t_*=16\,$min is
among the shortest ever published. The high-cadence, wide-field
observations that are the hallmark of KMTNet are the only way to
routinely capture such short crossings.  High-cadence resolution
of short caustic crossings will preferentially lead
to mass and distance measurements for the lens.  This is because the
short crossing time typically implies a nearby lens, which enables
the measurement of additional effects (bright lens and/or microlens
parallax).  When combined with the measured crossing time, these effects
can yield planet/host masses and distance.
}

\newcommand{\bdv}[1]{\mbox{\boldmath$#1$}}

\def\au{{\rm AU}}

\def\kms{{\rm km}\,{\rm s}^{-1}}
\def\masyr{{\rm mas}\,{\rm yr}^{-1}}
\def\muas{{\mu\rm as}}
\def\kpc{{\rm kpc}}

\def\mas{{\rm mas}}
\def\lim{{\rm lim}}

\def\rel{{\rm rel}}

\def\hel{{\rm hel}}
\def\geo{{\rm geo}}

\def\e{{\rm E}}

\def\bmu{{\bdv\mu}}

\def\ga{{>\atop \sim}}
\def\apj{{ApJ}}

\def\aap{{A\&A}}
\def\pasp{{PASP}}
\def\mnras{{MNRAS}}

\begin{document}
\jkashead 


\section{Introduction \label{sec:intro}}

The Korea Microlensing Telescope Network (KMTNet; \citealt{Kim:2016})
is a system of three 1.6m telescopes, each equipped with a
$4\,\rm deg^2$ 340-Mpixel camera, located in Chile, South Africa, and
Australia.
The system was designed to conduct a microlensing survey that would
detect Earth-mass planets without additional followup data.
While microlensing events typically last for tens of days, many of
their most interesting features occur on much shorter timescales,
 especially features that permit the detection and characterization
of planets, which have
typical perturbation times
\begin{equation}
t_p\sim \sqrt{m_p\over M_{\rm jup}}\,{\rm day} \sim 1.3 \sqrt{m_p\over M_{\oplus}} \rm hr,
\label{eqn:tp}
\end{equation}
where $m_p$ is the mass of the planet.

Most microlensing planets have been discovered by combining
survey and followup observations. In their original suggestion of
this approach, \citet{gouldloeb92} argued that
Galactic bulge fields should be monitored at relatively low cadence
(e.g., once per night).  This would enable them
to cover the widest possible area, which is necessary to detect
a large sample of microlensing events in spite of their low frequency
$\Gamma\sim 10^{-5} \rm yr^{-1}\,star^{-1}$, but would not generally
permit the detection of planets.
Hence, the survey teams would have to alert the community
to these events in real time, allowing followup networks to initiate
high-cadence monitoring of individual events.

The effectiveness of this survey+followup approach was greatly increased
by the recognition that the sensitivity of microlensing events to planets
is dramatically increased at high magnification \citep{griest98}.  This
permitted scarce observing resources to be focused on the rare subset of
high-mag events, with the observations concentrated on the brief
interval near peak.
A spectacular example
of the success of this approach was the discovery of the Sun/Jupiter/Saturn
analog OGLE-2006-BLG-109L \citep{ob06109,ob06109b}.
Despite its successes, this approach of combining survey and
followup observations is highly inefficient and allows only a small
subset of events to be observed with a high enough cadence to detect
planets. See \citet{gould10}
for a thorough review of this approach.

Even from the beginning, however, an alternate approach of survey-only
detections yielded a significant fraction of all planet detections.
For example, the very first microlensing planet OGLE-2003-BLG-235Lb
was discovered in this mode (although the planet characterization benefited
substantially by ``auto-followup'' by the MOA group)
\citep{ob03235}.  This event was unusual
in that the planetary perturbation lasted several days, which is
what enabled characterization given the surveys of that time.
However, these survey characteristics also permitted the survey-only
detection of a variety of other planetary events, including the
fleeting planetary signal in the high-magnification event MOA-2007-BLG-192
\citep{mb07192}.  Moreover, survey-only coverage yielded key light-curve
features even in some events for which the main features were discovered
in survey+followup mode, such as OGLE-2006-BLG-109L, mentioned above.

Nevertheless, as soon as the first microlensing planet was discovered,
it was clear that by augmenting the camera size, one could
cover the same area at much higher-cadence observations that would be capable of
detecting and characterizing planets without followup.  This insight
ultimately led to the MOA-II and OGLE-IV experiments, which were
commissioned in 2006 and 2010, respectively.

In 2004, Cheongho Han initiated discussions
with colleagues that led to the design of a more far-reaching variant
of this approach:
construct very wide-field cameras on the three southern continents.
By virtue of their
large fields of view, these telescopes could both monitor very wide areas and
do so at a rapid cadence (i.e., several times per hour).
Their geographic distribution would allow nearly continuous coverage of events.
Thus, such a survey would not only detect microlensing events but
would also be able to characterize planetary perturbations. Hence, this
approach allows all events to be monitored for planets and eliminates the need
for real-time followup observations.  A smaller scale variant of this
approach was achieved by combining the OGLE, MOA and Wise surveys
\citep{wise1,wise2}.

A natural consequence of this approach is that all features in the
microlensing light curve are automatically monitored at high
cadence. Aside from just detecting the planetary perturbations, it is
important to capture the caustic crossings because, if they are
resolved, they provide crucial additional information about the
lens. Caustic crossings occur in the great majority of published
planetary and binary events.  While there is some publication
bias hidden in this statistic, detailed simulations show that
even with systematics-free data, at least half of all recognizable
planetary events have caustic crossings \citep{Zhu:2014,Zhu:2015}.

The duration of the crossing allows one to determine the
normalized source size
\begin{equation}
\rho = {\theta_*\over \theta_\e} \,;
\ \theta_\e\equiv \sqrt{\kappa M \pi_\rel}\,;
\ \kappa\equiv {4 G\over c^2\au}\simeq 8.14\,{\mas\over M_\odot}
\label{eqn:rho}
\end{equation}
where $\theta_*$ is the angular radius of the source, $\theta_\e$ is the
Einstein radius, $M$ is the lens mass,
and $\pi_\rel = \au(D_L^{-1}-D_S^{-1})$ is the
lens-source relative parallax.
Since
$\theta_*$ is almost always measurable by standard techniques
\citep{ob03262}, this yields a measurement of $\theta_\e$ and
thus of the product $M\pi_\rel$, which can be a crucial constraint
in the interpretation of the lens.

Typical source sizes
are $\theta_*=0.6\,\muas$, and typical lens-source relative
proper motions are $\mu_\rel = 4\,\masyr$.  Hence, caustic crossing
times are similar in duration to Earth-mass planetary perturbations
\begin{equation}
t_{\rm cc} = t_*\sec\phi \,,
\
t_*\equiv {\theta_*\over\mu_\rel}
=1.3\,{\rm hr}\,{\theta_*\over 0.6\,\muas}
\biggl({\mu \over 4\,\mas/\rm yr}\biggr)^{-1},
\label{eqn:cc}
\end{equation}
where $t_*$ is the source self-crossing time, $t_{\rm cc}$ is the
caustic crossing time and $\phi$ is the angle of crossing. Therefore,
resolving a caustic crossing requires the same cadence as detecting
Earth-mass planets.

Although considerable effort is exerted by followup groups to monitor
caustic crossings of ``interesting'' events, in practice this is
difficult and suffers from many of the same problems as followup
searches for planets.
At least half of all caustic crossings (i.e., caustic entrances) are essentially impossible
to predict from the pre-crossing light curve, and a fair fraction
of the rest (i.e., caustic exits)
are quite difficult to predict.   In addition, many caustic exits
occur when they
are unobservable from a given location on Earth due to
sidereal time or weather.
Moreover, many events that turn out to be ``interesting''
are not recognized as such until after the time of the crossing.  This makes
high-cadence, round-the-clock surveys such as KMTNet a game-changer
for planetary and binary science.

Here we present a planetary microlensing event, OGLE-2015-BLG-0954,
that showcases KMTNet's ability to resolve unexpected caustic crossings
of exceptionally short duration.  We show that capturing the crossing
was crucial to the interpretation of this planetary system.

\section{KMTNet}

The three KMTNet telescopes were all commissioned in
the 2015 bulge season, taking their first scientific
data in February (South Africa), March (Chile), and
June (Australia).  During this commissioning season, they simultaneously
took scientific data and underwent various engineering tests and
adjustments.  While the system was only fully operational at the
end of the season, the data quality are generally quite high.
For example, these data played an important role in the characterization
of the massive-remnant binary candidate OGLE-2015-BLG-1285La,b
\citep{ob151285}.

The system is designed to monitor a large field at a rapid cadence
\citep{Kim:2016}.
The pixel scale
was optimized for microlensing at $0.4^{\prime\prime}$ to cover the widest
possible area while still being Nyquist sampled at typical good seeing
at the best site (Chile). The cadence was selected based on Equation~(\ref{eqn:tp}) which shows that it should
be of order 4--6$\,\rm hr^{-1}$ to enable approximately a dozen measurements
over the perturbation time of an Earth-mass planet $2t_{p,\oplus}\sim 3\,\rm hr$.
Indeed simulations by \citet{Henderson:2014} show that as one increases
the observing area (and correspondingly decreases the cadence) with
a KMTNet-like system, the number of planet detections of all types
increases at first.  However, at cadences of about $5\,\rm hr^{-1}$,
the number of Earth detections reaches a maximum while the detections of
more massive planets continues to rise.
Hence, the survey strategy adopted in 2015 was to observe 4 fields (so $16\,\rm deg^2$)
at about $6\,{\rm hr}^{-1}$.
In addition,
other fields were also observed at much lower cadence
in pursuit of other science objectives.

\section{OGLE-2015-BLG-0954's Unforeseen\\ Caustic Crossing \label{sec:unforseen}}

On 10 May 2015, the Optical Gravitational Lens Experiment (OGLE) Early Warning
System (EWS, \citealt{ews1,ews2}) alerted the microlensing community
to a new event, OGLE-2015-BLG-0954 at
$(\alpha,\delta)$=(18:00:44.24,$-28$:39:39.2), $(l,b)=(1.9,-2.7)$.
At HJD$^\prime = 7164.62$
(UT 02:45 22 May) the event began a caustic entrance, which
led to a jump of about 1 magnitude between the first and second
OGLE points of the night, which were separated by 105 minutes.
These were posted to the EWS website
 when OGLE next updated it, shortly after UT 15:01 22 May, i.e.
HJD$^\prime = 7165.12$, which happened to be almost exactly the
time of the end of the caustic exit HJD$^\prime = 7165.15$.  Hence, there was
essentially no time to organize observations of the caustic exit.
In any case, there were no such attempts of which we are aware.

\section{Observations \label{sec:kmtnetdata}}

OGLE-2015-BLG-0954 was observed only by the OGLE and KMTNet collaborations.
There were no followup observations.

OGLE observations were carried out with the 1.3m Warsaw Telescope at
Las Campanas, Chile.  The event lies in OGLE field BLG512, which means
that is normally observed at 20 minute cadence in $I$ band, with
occasional $V$-band observations to determine the source color.
The 105-minute gap reported above is therefore unusual.  There
are no observing-log notes on this hiatus, but given the conditions
at neighboring CTIO (see below), we suspect that the observer considered
the conditions at LCO as unacceptable.

KMTNet scientific observations commenced on 10 March, 21 February, and 20 June
in Chile (CTIO), South Africa (SAAO), and Australia (SSO), respectively.
Overall
the data from CTIO, SAAO, and SSO are of good science quality, but the commissioning observations
were impacted by a variety of engineering-related issues ranging from
breaks in the schedule to adjustment of the detector electronics.  In the case of the CTIO data there are two times when the
detector electronics changed substantially. We therefore
fit the CTIO light curves as three independent curves (to avoid introducing
artifacts into long term effects, particularly parallax), with breaks at
HJD$^\prime = 7130$ and HJD$^\prime = 7192$. We also note that because the SSO observations
began well after the caustic crossing, they do not contribute substantially
to the analysis, but are included for completeness.

The majority of data taken at CTIO and all the data taken at SAAO and SSO were
taken in $I$-band, with respectively 2132, 1350, and 449 points included
in this study.
$V$-band observations were made in Chile only, for the purpose of
determining the source color.  See Section~\ref{sec:cmd}.  The ratio of
$V$:$I$ observations in
Chile is about 1:5.  Only the second segment of Chile data were used
for this purpose because this is the only segment that has sufficient
flux variation to accurately constrain the source color.
There are 113 $V$-band points that could be matched to a corresponding
$I$-band point within 1.2 hours and so are used in this color determination.

Observations during the caustic crossing were taken in quite unstable
weather conditions.  The crossing lasted $2 t_{\rm cc} = 33\,$min.  Of the three
points taken during this crossing, one image could not be reduced and
the remaining two have formal errors that are respectively 3 and 2
times larger than typical errors at the same brightness during the
rest of the night.  Conditions recovered immediately after the caustic entrance.
KMTNet policy is to take observations  regardless of seeing and with up
to 6 magnitudes of extinction by clouds.  The great majority of observations
taken in poor conditions are ``useless'' in the sense that the information
content of neighboring points far exceeds that of these
poor-condition data.  However, as Lao Tzu\footnote{``Any man can make use of the
useful, but it takes a wise man to make use of the useless.''} foretold, these
``useless'' data can prove crucial.

The modeling was conducted based on reductions using
image subtraction \citep{albrow09,alard98}.
DoPhot \citep{dophot} reductions were used
to determine the source color and magnitude.

\section{Light Curve Model \label{sec:model}}

\begin{figure}[!t]
\centering
\includegraphics[trim=5mm 1mm 37mm 25mm, clip, width=84mm]{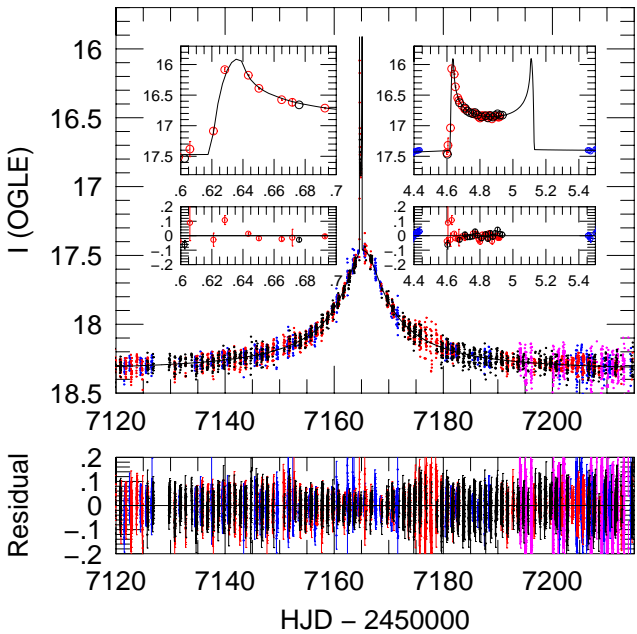}
\caption{Light curve and best-fit model for OGLE (black) and
KMTNet observations
of OGLE-2015-BLG-0954 with KMTNet data from Chile (red), South Africa (blue)
and Australia (magenta).  Insets show the caustic, which extends
from 7164.62 to 7165.15.  Error bars are omitted from the main
figure to avoid clutter but are shown in the residuals.
\label{fig:lc}}
\vskip1\baselineskip 
\end{figure}

Figure~\ref{fig:lc} shows the light-curve data from OGLE and the three
KMTNet observatories together with the best fitting model.
With the exception of a brief half-day eruption at peak, the
light curve looks qualitatively like a standard point lens event
that might adequately be parameterized by just three parameters,
$(t_0,u_0,t_\e)$, i.e., the time of lens-source closest approach,
the impact parameter (normalized to $\theta_\e$), and the
Einstein crossing time.  The caustics at peak imply that this
is a binary or planetary lens, which then requires three more
parameters $(s,q,\alpha)$, the separation of the two masses
(normalized to $\theta_\e$), their mass ratio, and their orientation
relative to the lens-source proper motion $\bmu_\rel$.  The fact that
the caustic entrance is resolved means that the normalized
source radius $\rho$ is also required for the model.  In addition,
source and blend fluxes $(f_s,f_b)$ must be fit for each independent
data set (3 from CTIO and 1 from each of the others).
We conduct
a broad search over $(s,q)$ geometries, allowing the other
five parameters to vary continuously from seed solutions that are
well sampled in $\alpha$.  We find only two local minima,
which are related by the well known close/wide binary degeneracy
\citep{griest98,dominik99}.  The parameters for these two
solutions are given in Table~\ref{tab:params}.

\begin{table}[t!]
\centering
\caption{Microlensing parameters for OGLE-2015-BLG-0954}
\begin{tabular}{llrr}
\toprule
Parameter & Unit & Close & Wide \\
\midrule
$\chi^2$        &                & 12909.1 & 12908.6\\
 /dof           &                & / 12916& / 12916\\
 \midrule
$t_0 - 7100$        &day             &65.223 & 65.220 \\
                    &                &0.009 & 0.009 \\
 \midrule
$u_0$               &                & $-$0.0568 & $-$0.0580 \\
                    &                & 0.0017 & 0.0021 \\
 \midrule
$t_{\rm E}$         &day               &37.53 & 36.96 \\
                    &                & 0.87 &  1.10 \\
 \midrule
$s$                 &                &0.7982 & 1.3522 \\
                    &                &0.0033 & 0.0063 \\
 \midrule
$q$                 &                &0.01098 & 0.01194 \\
                    &                &0.00028 & 0.00043 \\
 \midrule
$\alpha$            &radian          & 4.5462 & 4.5890 \\
                    &                & 0.0044 & 0.0044 \\
 \midrule
$\rho$              &$10^{-4}$        & 2.96 & 3.03 \\
                    &                & 0.11 & 0.13 \\
\bottomrule
\end{tabular}
\label{tab:params}
\end{table}

There are two major striking features of these solutions.
First, the planet/star mass ratio is 0.011 (or 0.012), i.e.,
about 11 or 12 times that of Jupiter/Sun.  Second, the self-crossing
time is extremely quick,
\begin{equation}
t_* = \rho t_\e = 16.0\pm 0.5\,{\rm min}.
\label{eqn:tstar}
\end{equation}
Comparison with Equation~(\ref{eqn:cc}) shows that either the
source must be extremely small or $\mu_\rel$ must be very high.

\section{Color-Magnitude Diagram \label{sec:cmd}}

\begin{figure}[!t]
\centering
\includegraphics[trim=5mm 10mm 35mm 25mm, clip, width=84mm]{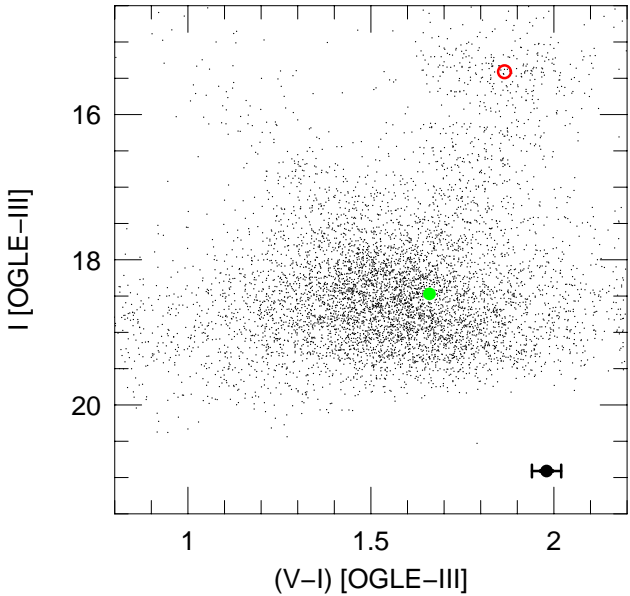}
\caption{Calibrated color magnitude diagram showing the
microlensed source (black) and the blended light (green)
relative to the centroid of the red clump (red).  The
field stars are taken from the OGLE-III catalog \citep{ogle-iii}.
The source color and magnitude are derived in instrumental
systems, and then transformed to the OGLE-III system
using field stars.  The blended light is derived by
subtracting the source $(V,I)$ fluxes from the $(V,I)$
fluxes of the baseline object found in the OGLE-III catalog.
\label{fig:cmd}}
\end{figure}

Figure~\ref{fig:cmd} is a calibrated color-magnitude diagram
(CMD) showing field stars within $90^{\prime\prime}$ of the lens from
the OGLE-III catalog \citep{ogle-iii} together with the source
position determined from the model presented in Section~\ref{sec:model},
as well as the blend.

To construct this diagram, we make two, basically independent measurements,
using KMTNet and OGLE data, respectively

For the KMTNet measurement, we first measure the source in
an instrumental system based on DoPhot photometry using regression
for the $(V-I)$ color and a fit to the best model for the $I$ magnitude,
respectively.  We then put these on the OGLE-III system via a transformation
derived from common field stars.  We find
$((V-I),I)_s = (2.01\pm 0.05, 20.92\pm 0.03)$, where the error
in the color is due to the scatter in the photometric
data, while the error in the magnitude is primarily due to a highly correlated
$\sim 3\%$ error in the parameters $(f_s,u_0,t_\e,q,\rho)$.
See Table~\ref{tab:params} and \citet{mb11293}.

We repeat this procedure using OGLE-IV light-curve data, which is calibrated
in $I$ but not $V$.  We find
$((V-I),I)_s = (1.91\pm 0.07, 20.90\pm 0.03)$.  Noting that the errors
in the color are independent but those in the magnitude are not, we combine
the two measurements to obtain  $((V-I),I)_s = (1.98\pm 0.04, 20.91\pm 0.03)$.

To derive the calibrated blend (i.e., non-source)
light, we subtract the fluxes corresponding to these magnitudes
from those of the OGLE-III star at this position, with
$((V-I),I)_{\rm base}=(1.69,18.36)$ to obtain
$((V-I),I)_b=(1.66,18.47)$. The errors in these quantities are essentially
the same as those of the baseline flux because the blend fluxes are almost the
same as the baseline fluxes.  We estimate these as 0.07 and 0.02 magnitudes
respectively.

Assuming that the source is behind the same amount of
dust as the clump, we can determine its intrinsic color and magnitude
and use those values to determine $\theta_\e$.
First, we measure the position of the
red clump: $((V-I),I)_{\rm cl}=(1.87,15.41)$.
Given the intrinsic
clump position $((V-I),I)_{\rm 0,cl}=(1.06,14.38)$
\citep{bensby13,nataf13}, we then derive
$((V-I),I)_{0,s}=(1.17,19.88)$. Then using the $VIK$ color-color
relations of \citet{bb88} and the color-surface brightness relations
of \citet{kervella04}, we derive an angular source radius,
\begin{equation}
\theta_* = 0.56\,\muas.
\label{eqn:thetastar}
\end{equation}
The resulting angular size of the Einstein ring and
lens-source relative proper motion are
\begin{equation}
\theta_\e = \frac{\theta_*}{\rho} = 1.89\,\mas
\,;\quad
\mu_\rel = \frac{\theta_\e}{t_\e} = 18.4{\mas\over \rm yr}\,.
\label{eqn:thetae}
\end{equation}
All three quantities have the same 9\% error, which
we calculate as follows.  The 3\% modeling flux error contributes
half \citep{ob08279}, i.e., 1.5\%.  Centroiding the clump in $I$
contributes 4\%.  There is a 0.04 mag measurement error in the color
and another 0.06 mag error
in deriving the surface temperature from the microlensing color
\citep{bensby13}, which together propagate to a 6\% error.
Finally, there is a 5\% error from propagating the
scatter in the $VIK$ relations.
\\[-3.25pt plus1pt minus1pt] ~~

\section{The Nature of the Lens \label{sec:lens}}

\subsection{A Nearby Lens}

The measured $\theta_\e$, blended light, and proper motion strongly
imply a nearby lens.

\subsubsection{$\theta_\e$ and the CMD \label{sec:uplimcmd}}

We can conclude with near certainty that lens is in
the near disk from the large Einstein radius combined with
limits on the light from the lens, which cannot be brighter than
the observed blended light.  The $\theta_\e$ measurement provides a
constraint from $\theta_\e^2 = \kappa M\pi_\rel$, i.e.,
\begin{equation}
M = (1.0\pm 0.18)M_\odot\biggl({\pi_\rel\over 0.44\,\mas}\biggr)^{-1} .
\label{eqn:mpirel}
\end{equation}
Suppose that the lens were a $M=1\,M_\odot$ star, and therefore at 1.75
kpc. Even if it were behind the same amount of dust as the clump
($A_I=1.05$, \citealt{nataf13}) and were somewhat subluminous at
$M_I=4.5$, it would be $I\sim 16.75$, about 1.7 mag brighter than
the observed blend.  In addition, it would be somewhat bluer.
Essentially the same argument applies to any host $M\gtrsim 0.25\,M_\odot$
(i.e, $\pi_\rel=1.76\,\mas$, $M_I\sim 9$ $\Rightarrow I=18.67$).
That is, it results in a solution that contradicts the observed blend.

Note that at the limit $M=0.25 M_\odot$, the photometric constraint from
the blended light would still not be truly satisfied. That is,
although the lens would contribute less than the observed I-band
light, the color would still be wrong. Such a star would contribute
only about 15\% of the observed V-band light, meaning the other 85\%
would then have to be supplied by another star. Of course, there could
be such an object, but it would itself contribute substantial I-band
light. When combined with the lens light, the total light would then
exceed the limit unless the lens were yet fainter than we have just
supposed.

On the other hand, if $\theta_\e$ were $2\sigma$ smaller, this would
allow for smaller $\pi_\rel$ at fixed mass, and hence, a more
distant lens. This consequent 0.4 mag reduction in flux from the
lens would provide the necessary margin to accommodate light from an
additional star.

Taking all of these considerations into account, we therefore regard
$M < 0.25 M_\odot$ as a conservative photometric upper limit on the
host mass. This gives a corresponding upper limit on the lens distance
of 0.5 kpc. This leads to the conclusion that the lens is a very-low
mass, nearby object and may even be a brown dwarf.

The only exception to this line of reasoning would be stellar remnants,
either white dwarfs, neutron stars, or black holes.  These are
generally believed to have few planets so we do not further pursue
this possibility here.  As we discuss in Section~\ref{sec:resolve},
if future observations show that the lens is not contributing to the blend light
then this question should be revisited.

\subsubsection{Lens-Source Relative Proper Motion}

The lens-source relative proper motion,
$\mu_\rel =18\,\masyr$ (Equation \ref{eqn:thetae}),
also suggests a nearby lens.  This value is much higher than
for typical bulge lenses $(\sim 4\,\masyr)$ or far-disk
lenses $(\sim 7\,\masyr)$, which are dominated by internal
stellar motions in the bulge and the flat rotation curve of the
Galaxy, respectively.  However, the proper motions of very nearby lenses
are dominated by their peculiar motion, $v_{\rm pec}\sim {\cal O}(50\,\kms)$,
i.e., $\mu_\rel \rightarrow v_{\rm pec}/D_L$.  This can be arbitrarily
large for sufficiently nearby lenses.  For example, the extremely
nearby lens OGLE-2007-BLG-224 $(D_L= 0.5\,\kpc)$ had a record
proper motion of $\mu_\rel=48\,\masyr$ \citep{ob07224}.
While the observed proper motion does not
definitively rule out either the bulge-lens or far-disk lens
scenarios, it is most easily explained by a nearby lens.

\subsection{Lens Properties \label{sec:properties}}

Given our conclusion that the lens is nearby, we can estimate its mass and
distance based on the velocity distribution of stars in the nearby disk. The
projected transverse velocity $\tilde v\equiv \au\mu_\rel/\pi_\rel$ is
typically 50 km s$^{-1}$, and the lens mass is related to $\tilde v$ by
\begin{equation}
M = {c^2\over 4 G}\tilde v t_\e\theta_\e \rightarrow
0.25\,M_\odot{\tilde v\over 50\,\kms},
\label{eqn:mvtilde}
\end{equation}
where we have substituted in the best fit values for $\theta_\e$ and $t_\e$
in the last step.  Similarly,
\begin{equation}
\pi_\rel = {\au\mu_\rel\over\tilde v}\rightarrow
1.7\,\mas\biggl({\tilde v\over 50\,\kms}\biggr)^{-1}
\label{eqn:pirelvtilde}
\end{equation}
Since, the rms projected velocity of nearby stars is $\tilde v\sim 50\,\kms$,
these are plausible first estimates for the lens mass and relative parallax.
They are roughly consistent with the upper limits derived in
Section~\ref{sec:uplimcmd}.

However, we can also make a purely kinematic estimate ignoring the
flux constraints.  This is appropriate either to allow for dark
(or very dim hosts) such as white dwarfs, or to allow for larger-than-expected
errors in $\theta_\e$ (which entered these constraints), due
to either large statistical fluctuations or to unrecognized systematic errors.

To do so, we must also weight our results to account for the
transformation between microlensing parameters and the physical
parameters of the system.  This accounts for the underlying bias
of the observations
toward more distant and thus (given that
$M\pi_\rel=\theta_\e^2/\kappa$ is measured) more massive lenses.

For fixed (i.e., measured) $\theta_\e$ and $\mu_\rel$, the
probability is weighted by the volume element $(d D_L\, D_L^2$) as
well as the stellar density variation (which in the present case is minor).
Approximating the projected velocity distribution\footnote{We estimate
$2\sigma^2 = \sigma_\phi^2 + \sigma_z^2 + v_{\oplus,\perp}^2 +
[(D_L/D_S)v_{\rm rot}]^2$ where $(\sigma_\phi,\sigma_z)=(33,18)\,\kms$
are the Galactic velocity dispersions of the disk in the rotation and
vertical directions,
$v_{\oplus,\perp}=25\,\kms$ is Earth's speed relative to the LSR at the
peak of the event, and $(D_L/D_S)v_{\rm rot}= \kms$ is the projection
of Galactic rotation.  This would yield $\sqrt{2}\sigma=48\,\kms$, which
we round to 50$\,\kms$.}  as a Gaussian
with 1-D dispersion $\sigma = 50\,\kms/\sqrt{2}$, one may write,
\begin{equation}
\langle M \rangle = {\int d D_L [D_L(\tilde v)]^2 M(\tilde v)
\tilde v\exp(-\tilde v^2/2\sigma^2)\over
\int d D_L [D_L(\tilde v)]^2 \tilde v\exp(-\tilde v^2/2\sigma^2)},
\label{eqn:meanm}
\end{equation}
where $M$ and $D_L$ are regarded as implicit functions of $\tilde v$
through Equations~(\ref{eqn:mvtilde}) and (\ref{eqn:pirelvtilde}),
respectively.  Then using the near-field approximation $D_L\rightarrow \au/\pi_\rel$, one may easily derive
\begin{equation}
M = {c^2\over 4 G}\sqrt{2}\sigma\biggl({3\sqrt{\pi}\over4}\biggr)
t_\e\theta_\e \rightarrow
0.33\pm 0.12\,M_\odot
\label{eqn:mvtildeest}
\end{equation}
and similarly for $\pi_\rel$,
\begin{equation}
\pi_\rel = {\au\mu_\rel\over\sqrt{2}\sigma}\biggl({\sqrt{\pi}\over 2}\biggr)
\rightarrow
1.5\pm 0.6\,\mas.
\label{eqn:pirelvtildeest}
\end{equation}

These estimates are consistent at the 1-$\sigma$ level with the
host-mass limit $M<0.25\,M_\odot$ that was derived in
Section~\ref{sec:uplimcmd}
and also at the 1.5-$\sigma$ level
with typical white-dwarf masses.  In Section~\ref{sec:resolve},
we discuss how this ambiguity can be resolved.  Until that time,
we adopt the above kinematic estimates, which straddles both cases.
Then,
combining Equation~(\ref{eqn:mvtildeest}) with the planet-star mass ratio
$q=0.0115$ (average of the two solutions) yields an estimated planet
mass,
\begin{equation}
m_{\rm planet} = 3.9\pm 1.4\,M_{\rm jup} .
\label{eqn:planetmass}
\end{equation}
When high-resolution imaging resolves the nature of the host
(Section~\ref{sec:resolve}), this estimate can be made more precise.

Because the distance uncertainty is roughly 30\% and the two solutions
(wide and close) are roughly equally likely, the estimates of the
projected separation from these two solutions overlap.  We therefore
estimate
\begin{equation}
a_\perp = D_L\theta_\e s \rightarrow 1.2 \pm 0.6\,\au .
\label{eqn:planetsep}
\end{equation}
Note that while some of this large uncertainty can be removed by resolving
the host, the dichotomy between close and wide solutions will remain.
\\[-3.25pt plus1pt minus1pt] ~~

\section{What is the Nature of the Blended Light? \label{sec:blend}}

\subsection{Astrometry \label{sec:ast}}

To probe the relation of the blended light to the event, we first
determine whether or not it is coincident with the position of the
lens by measuring the centroid of the source-blend combined light. We
compare
the apparent position of the source-blend combined light when the
source is highly magnified on HJD$^\prime \sim 7187.7$ (when the source
accounts for 65\% of the total light) to its position near baseline
(when it accounts for about 13\%).  We conduct two independent
measurements, using KMTNet and OGLE data, respectively, employing
somewhat different procedures for the two cases.

For the KMTNet data, we consider 33 CTIO images
with seeing FWHM $<1.35^{\prime\prime}$, including 7 from the above mentioned
highly magnified night and 26 that are at least 25 days from peak (so
magnified $A<1.8$).  Because the baseline object is relatively isolated
(with only one relevant near-neighbor, which is at $1.4^{\prime\prime}$),
we fit directly to the flux counts of individual pixels in the
vicinity of the lens, taking account of this neighbor.
We find that the offset between these two groups (peak minus baseline)
is $-60\pm 25\,\mas$ and $0\pm 27\,\mas$ in the East and North directions.
This implies separations between the source and blend of
$-120\pm 50\,\mas$ and $0\pm 54\,\mas$, respectively.

For the OGLE data we compare the apparent position of the difference
image (isolated PSF obtained from magnified image minus template)
near maximum to the DoPhot position of the apparent source (really source+blend) at baseline
(mean epoch 2010.8).  We find offsets of
$-80\pm 25\,\mas$ and $+105\pm 25\,\mas$ in the East and North directions,
which corresponds to separations between the source and blend of
$-90\pm 30\,\mas$ and $+120\pm 30\,\mas$, respectively.

Naively, the OGLE and KMTNet measurements are consistent at much better than
$1\,\sigma$ in the East direction but disagree at $2\,\sigma$ in
the North direction.  This itself is hardly unusual and would
be expected $\exp(-2^2/2)=14\%$ of the time.  Moreover, we must
bear in mind that these two measurements are not necessarily measuring
the same thing.  After some algebra, one may show that if the blend is associated with the lens (being either
the lens itself or a companion to the lens), then one expects the OGLE and KMTNet measurements to differ by
$\vec\mu\Delta t\sim 85\,\mas$, where $\mu=18\,\masyr$ in some unknown direction and  $\Delta t\sim 4.7\,$yr.
Hence, the two measurements may be more consistent than first appears.

In brief, both the OGLE and KMTNet measurements give a consistent
measurement of the separation between they source and the blend. They
are both consistent at the 2-$\sigma$ level with zero separation between
the source and blend. At the same time, if the source and blend are
not coincident, these measurements place an upper limit on their
separation of 200 mas.  Hence, there is no clear evidence that the
blended light is displaced from the source. We discuss how to improve
this constraint in Section \ref{sec:resolve}.

\subsection{Four Possible Origins of Blended Light}

\subsubsection{Blend as Unrelated Star}

There are 3500 stars brighter than the blend within $90^{\prime\prime}$
of the lens.  Hence, the chance of one of these being projected within
$200\,\mas$ is $3500\times (0.2/90)^2 = 1.7\%$.  While this probability
is small, such random projection is not implausible.

\subsubsection{Blend as Companion to the Source}

Another possibility is that the blend is a subgiant companion to the
source.  Roughly 15\% of G dwarfs have companions $M_{\rm comp}/M_G>0.8$
and with semimajor axes $a\lesssim 200\,\mas\times 8\,\kpc = 1600\,\au$
\citep{Raghavan:2010}.  Since reddish subgiants $(V-I)_0\ga 0.9$
live about 10 times shorter than such upper main-sequence stars,
the probability of such a companion is about 1.5\%, i.e., similar
to the first scenario.

\subsubsection{Blend as the Host}

The remaining possibility is that the blend is part of the
lens system, either the host itself or a companion to it
that is too far separated to influence the magnification.

It is difficult to come up with a plausible  host that explains the
blended light for reasons that are closely related to the fact
that the blended light places upper limits on the host light,
as discussed in Section~\ref{sec:uplimcmd}.
For example, if the blend were
behind all the dust, then its color would imply a late G dwarf of mass
$M\sim0.9\,M_\odot$.  Identifying this star with the host would then imply,
via Equation~(\ref{eqn:mpirel}), $D_L=1.6\,\kpc$, and so
$M_I=6.4$.   This is substantially too dim for typical late G dwarfs.
On the other hand, if
the lens were in front of all the dust, its color would imply a late K dwarf,
$M\sim 0.6\,M_\odot$ and so (assuming it were the host)
$\pi_\rel = \theta_\e^2/\kappa M=0.73\,\mas$,
i.e., $D_L\sim 1.2\,\kpc$ and so
$M_I=8.1$.   Again, this is substantially too dim for late K dwarfs.

\subsubsection{Blend as Companion to the Host}

Another possibility is that the blended light is a companion to the
host.  However, it is also not easy to arrange for this scenario.
For example, if the blend is in front of all the dust, then its
color implies a late K dwarf, so $M_I\sim 6$.  Then its apparent magnitude
implies $D_L=3\,\kpc$, and so $\pi_\rel\sim 0.2\,\mas$. But that implies
the host would be $M\sim 2.5\,M_\odot$, which would produce an obvious flux contribution
(unless it were a dark remnant).

Hence, although the probabilities for a random interloper (Section 8.2.1)
and a companion to the source are modest (Section 8.2.2), these are the most
likely scenarios.

\subsubsection{Blend as Combination of Two or More Stars}

Finally, of course, it is possible that the blended light
is due to two or more stars.  If both were drawn from the first
two possibilities above (source companion or random interloper)
then the prior probability would be roughly the square of the
probabilities calculated there, i.e., $\sim 3\times 10^{-4}$.
However, it is also possible that star light from one of these two
possibilities is combined with light from the lens system (host or
companion).  The large range of such possible combinations
would then imply that high-resolution imaging would be required to
extract a unique interpretation. We discuss this in the next section.

\subsection{Resolution \label{sec:resolve}}

Several types of information can be assembled on various
timescales to resolve the nature of the host.  First, it should
be possible to immediately measure the heliocentric proper motion of the
baseline-object using OGLE-III data (e.g., \citealt{mb11262b}).
If the proper motion is high, then it is very likely that the
blended light is either the host itself or that it is a companion
to the host. Note that in interpreting such a measurement
one must account for the difference between heliocentric and
geocentric proper motion,
\begin{equation}
\bmu_\hel -\bmu_\geo = {\bf v}_{\oplus,\perp}{\pi_\rel\over\au}
= (0.29,5.41){\pi_\rel\over \rm yr}
\label{eqn:deltamu}
\end{equation}
where we have substituted in the value of ${\bf v}_{\oplus,\perp}$,
Earth's instantaneous projected velocity at the peak of the event,
in (north,east) coordinates.  For example, if the lens were at
$\pi_\rel = 2\,\mas$, this difference would be more than $10\,\masyr$.

High resolution imaging of the lens system would yield more definitive
information.
Immediate high resolution imaging could determine whether the blend is
significantly offset from the source with much better constraints than
we found in Section \ref{sec:ast}.  If the blend is indeed separated
by $\Delta\theta = 120\,\mas$, it would easily be resolved.
If the high-resolution imaging is done in the optical (e.g., with
the {\it Hubble Space Telescope}) and if the blend is separated
from the source, then one could subtract the source flux
(known from the microlensing solution) from the combined light at the
position of the source, in order to measure the flux
of the lens (or at least put much better limits on it).
Since the lens is moving at
$\mu_\rel=18\,\masyr$, it could be resolved within 4 years
\citep{Batista:2015,Bennett:2015}.  Moreover, if the blend is now
separated from the source, combined first plus second epoch of
high-resolution imaging could determine whether the blend has the same
proper motion as the host and so is a companion to it.

The final scenario is that the host is a white dwarf of, e.g., $M=0.6\,M_\odot$, placing it
at $D_L\sim 1.1\,\kpc$.  Depending on its age and the instrument with
which it is observed, it will likely be either very faint or invisible.
However, in either case this would be an important discovery.  And in the
latter case, deeper observations could then be undertaken to
discriminate between brown dwarf and white dwarf scenarios.

\section{Conclusion \label{sec:conclude}}

High-cadence $\Gamma = 6\,{\rm hr}^{-1}$ observations by KMTNet
resolved the extremely short $2t_{\rm cc} = 33\,$minute caustic
crossing of the planetary event OGLE-2015-BLG-0954.  This
proved crucial to both the measurement of the Einstein radius
($\theta_\e=1.7\,\mas$) and to the arguments that the lens must
be in the near disk (from the high proper motion $\mu_\rel=18\,\masyr$).
Hence, it was crucial both to the mass estimates of the
host ($0.33\pm 0.12\,M_\odot$)
and planet ($3.9\pm 1.4\,M_{\rm jup}$), and to the recognition that
the lens and source will be separately resolvable in just a few years.

While such short caustic crossings are relatively rare, they are more likely
to lead to complete mass and distance solutions than
for typical events. This is because the most likely reason for the short
crossing time is very high lens-source proper motion, which in turn is most
likely due to a very nearby lens. It is easy to measure the flux of such
nearby lenses if they are of relatively high mass. On the other hand if they
are low mass, their microlens parallax \citep{gould92}
\begin{equation}
\pi_\e = {\pi_\rel\over\theta_\e} = \sqrt{\pi_\rel\over \kappa M}
\end{equation}
will be large and hence more easily
measurable.  Either of these effects can provide the
second parameter needed for a complete solution, provided that the Einstein
radius has been measured by resolving the caustic crossing.

In the case of OGLE-2015-BLG-0954, we were able to constrain the
mass of the lens from the limit on lens light. No parallax effects
were detected in the light curve.  Hence, we could not obtain a definitive
measurement.  Instead we estimated the lens mass and distance based
on kinematic arguments from its high proper motion.  These are consistent
with independent arguments derived from upper limits on lens light.
They are also consistent with a white dwarf host, although this would
be more surprising.
The nature of the lens can be completely resolved in a few years based on high
resolution imaging.

\acknowledgments

This research has made use of the KMTNet system operated by KASI and
the data were obtained at three host sites of CTIO in Chile, SAAO in South
Africa, and SSO in Australia.

This work was supported by KASI (Korea Astronomy and Space Science
Institute) grant 2016-1-832-01.

Work by J. C. Y. was performed under contract with
the California Institute of Technology
(Caltech)/Jet Propulsion Laboratory (JPL)
funded by NASA through the Sagan
Fellowship Program executed by the NASA Exoplanet Science
Institute.

The OGLE team thanks Profs. M. Kubiak and G. Pietrzy{\'n}ski, former
members of the OGLE team, for their contribution to the collection of
the OGLE photometric data over the past years.

The OGLE project has received funding from the National Science Centre,
Poland, grant MAESTRO 2014/14/A/ST9/00121 to A. U.

A. G. acknowledges support from NSF grant AST-1516842.



\end{document}